\documentclass{aa}
\usepackage{graphicx}
\def\simlt{\lower.5ex\hbox{$\; \buildrel < \over \sim \;$}}
\def\simgt{\lower.5ex\hbox{$\; \buildrel > \over \sim \;$}}

\def\taurus2{{\sc Taurus-2}}

\def\photunits{photons\,s$^{-1}$\,cm$^{-2}$}

\def\ergs2band{erg\,s$^{-1}$\,cm$^{-2}$\,band$^{-1}$}

\def\as{$''$}

\def\aspix{$''$\,pix$^{-1}$}

\def\MpcPer3{Mpc$^{-3}$}
\def\Mpc3{Mpc$^{3}$}

\def\cmPerSec{cm\,s$^{-1}$}

\def\masyr{mas\,yr$^{-1}$}

\def\Niib{{[{{\sc N}\,{\sc ii}}]~$\lambda$6548}\/}
\def\Niir{{[{{\sc N}\,{\sc ii}}]~$\lambda$6584}\/}
\def\NII{{[{{\sc N}\,{\sc ii}}]}\/}
\def\HI{{{\sc H}\,{\sc i}}\/}
\def\Ha{{H$\alpha$}\/}

\begin{document}
   \title{Discovery of an optical bow-shock around pulsar B0740$-$28\thanks{
Based on observations made with ESO Telescopes at the La Silla
Observatory (Programme 66.D--0046) and under the AAT Service Mode Programme}}

   \author{D.~Heath Jones\inst{1,2}, Ben W.~Stappers\inst{3,4} \and
	Bryan M.~Gaensler\inst{5}}
   \offprints{D.~H.~Jones, ({\tt hjones@eso.org})}

   \institute{European Southern Observatory, Casilla 19001, Santiago 19, Chile, \email{hjones@eso.org}
\and
Observatorio Cerro Cal\'{a}n, Departamento de Astronom\'{i}a,
Universidad de Chile, Casilla 36-D, Santiago, Chile,
\and
Stichting ASTRON, 7990 Dwingeloo, The Netherlands,
 	\email{stappers@astron.nl}
\and
Sterrenkundig Instituut `Anton Pannekoek', 1098 SJ Amsterdam, The Netherlands
\and
Harvard-Smithsonian Center for Astrophysics, 60 Garden Street, Cambridge MA 02138, USA
}

   \date{Received ; accepted }

   \abstract{
We report the discovery of a faint H$\alpha$ pulsar wind
nebula (PWN) powered by the radio pulsar B0740$-$28. The characteristic
bow-shock morphology of the PWN implies a direction of motion
consistent with the previously measured velocity vector for the
pulsar.  The PWN has a flux density more than an order of magnitude
lower than for the PWNe seen around other pulsars, but, for a distance
2~kpc, it is consistent with propagation through a medium of atomic density
n$_{\rm H} \sim 0.25$~cm$^{-3}$, and neutral fraction of 1\%. The
morphology of the PWN in the area close to the pulsar is distinct from
that in downstream regions, as is also seen for the PWN powered by
PSR~B2224$+$65. In particular, the PWN associated with PSR~B0740--28 
appears to close at its rear, 
suggesting that the pulsar has recently passed through a
transition from low density to high density ambient gas.
The faintness of this source underscores that deep
searches are needed to find further examples of optical pulsar nebulae.

}

\authorrunning{Jones, Stappers \& Gaensler}
\titlerunning{Discovery of an Optical Bow-Shock Nebula}

   \maketitle
%

\section{Introduction}

Pulsar wind nebulae (PWN) around high-velocity pulsars provide a primary
insight into the interaction between a pulsar and its environment.
Specifically, optical observations of such nebulae provide important
information on pulsar velocities, and on the density, temperature and
composition of the ambient medium. However only three pulsars are known to
power optical bow-shock PWNe: B2224$+$65 (the `Guitar Nebula', Cordes, 
et al.\ \cite{cordes93}), and the two millisecond pulsars (MSPs)
B1957$+$20 (Kulkarni \& Hester \cite{kulkarni88}) 
and J0437$-$4715 (Bell et al.\ \cite{bell95}).  
All three of these pulsars have high spin-down luminosities and/or high space
velocities. However, these pulsars differ markedly in their spin-periods,
ages and magnetic field strengths, highlighting
the variety of pulsar winds which can be probed by
these sources. The nebula associated with the
neutron star RX~J1856.5$-$3754 
further exemplifies the variety of neutron stars
known to power such nebulae (van Kerkwijk \& Kulkarni \cite{vankerkwijk01}).
We have therefore initiated a search for optical
bow-shocks around other pulsars in order to characterise the
properties of the associated pulsars, pulsar winds and ambient
environments.

\begin{table} 
\begin{center} 
\caption{Properties of  PSR B0740$-$28}
\label{physParams}
\begin{tabular}{lcr} 
\hline \hline
Spin Period$^a$		&	P (s)	& 0.1667 \\
Period Derivative$^a$	&	$\dot{P}$ & 1.68$\times10^{-16}$ \\
Spin-down Luminosity$^a$	&	$\dot{E}$ (ergs s$^{-1}$) & 1.4$\times10^{35}$ \\
Dispersion Measure Distance$^b$	&	d$_{\rm DM}$ (kpc) & 1.9$^{+0.4}_{-0.3}$ \\
H{\sc i} Distance$^c$	& d$_{\rm HI}$ (kpc) & 1.4 - 7.7 \\
Proper Motion$^d$       & $\mu$ (\masyr) & $29 \pm 1$ \\ 
Position Angle$^d$	& $\theta$ ($^\circ$) & $269 \pm 1$\\
\hline
\end{tabular} 
\end{center} 
$(a)$ Arzoumanian et al.\ (\cite{arzoumanian94}). 
$(b)$ Taylor \& Cordes (\cite{taylor93}). 
$(c)$ Koribalski et al.\ (\cite{koribalski95}). 
$(d)$ Bailes et al.\ (\cite{bailes90}). 
\\
\end{table} 

During the first two nights of this programme (January 4 and 5, 2001), we
discovered an optical bow-shock nebula around the radio pulsar
B0740$-$28. Discovered at Bologna (Salter \cite{salter70}), this pulsar
was subsequently shown to be rapidly spinning down (McCulloch et al.\
\cite{mcculloch73}).  Its correspondingly high spin-down luminosity,
$\dot{E}=4\pi I\dot{P}/P^3$, (where $I=10^{45}I_{45}$~g~cm$^{2}$ is the
pulsar moment of inertia; see Table \ref{physParams}), but likely moderate
distance and high transverse velocity make it a promising target
for powering an observable \Ha\ nebula.  We note that the maximum distance
derived from \HI\ measurements given in Table \ref{physParams} is large, but
in Section~5 we argue for a distance closer to that derived from the
dispersion measure. See Chatterjee \& Cordes (\cite{chatterjee02}) for a
recent review.

\begin{table} 
\begin{center} 
\caption{Log of Observations for the PSR B0740$-$28 field.\label{table2}}
\label{observations}
\begin{tabular}{lcc} 
\hline \hline
Date 			& 4, 5 January 2001 	& 14 April 2001\\
Instrument		& SUSI-2/NTT		& TTF$^a$/AAT \\
Pixel Scale (\aspix) 	& 0.161 		& 0.373 \\
Filter$^b$ (nm)		& 656/7			& 656/2 \\
Exposure Time (s) 	& 6480			& 2700 \\
Transparency		& photometric		& photometric \\
Mean Seeing (\as)	& $\sim 0.9$		& $\sim 1.4$\\
Continuum Frame		& None			& 600 s in $R1^c$\\
\hline
\end{tabular}
\end{center} 
$(a)$ using the blue Fabry-Perot and $R0$ (668/21) filter.\\
$(b)$ central wavelength and FWHM in nanometers. \\
$(c)$ TTF $R1$ filter (707/26). \\
\end{table} 

\section{Observations and Reduction}

\begin{figure*}
   \centering
   \includegraphics[width=6cm]{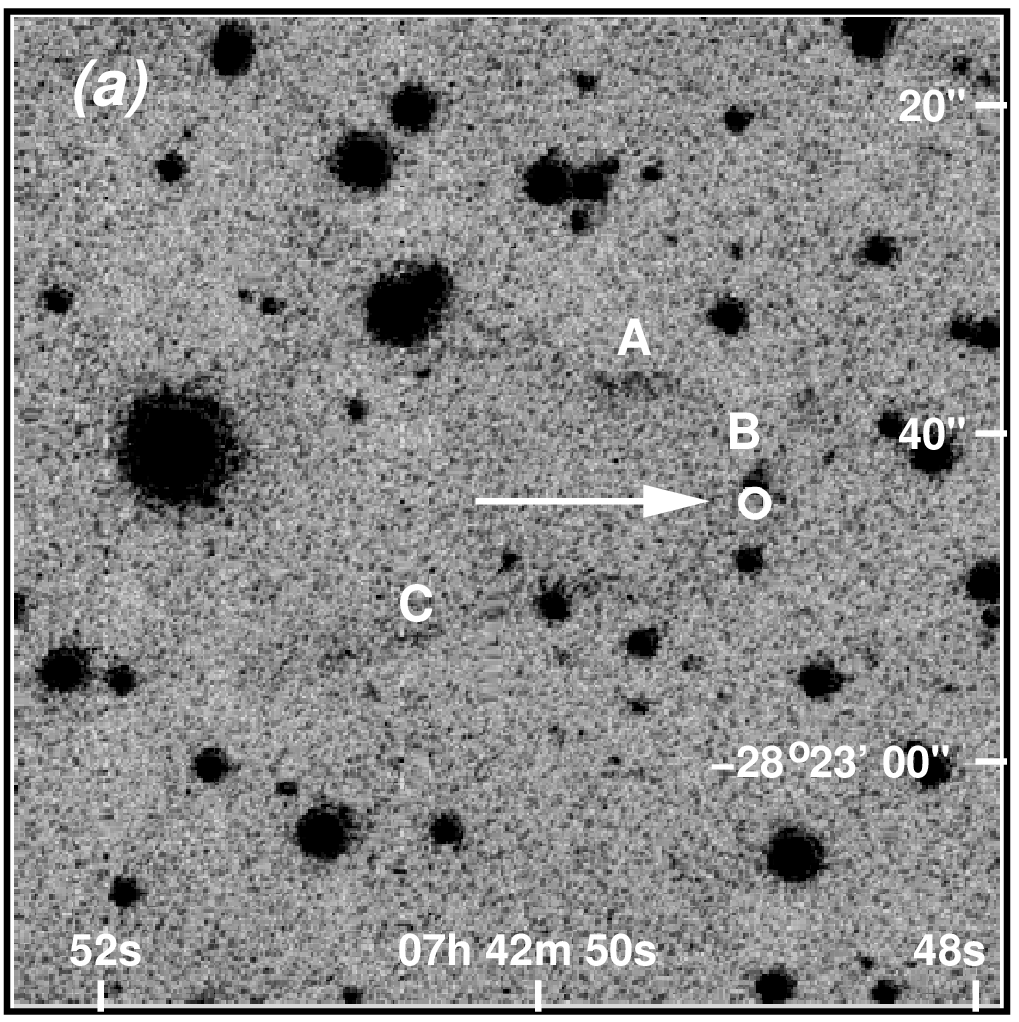}
   \includegraphics[width=6cm]{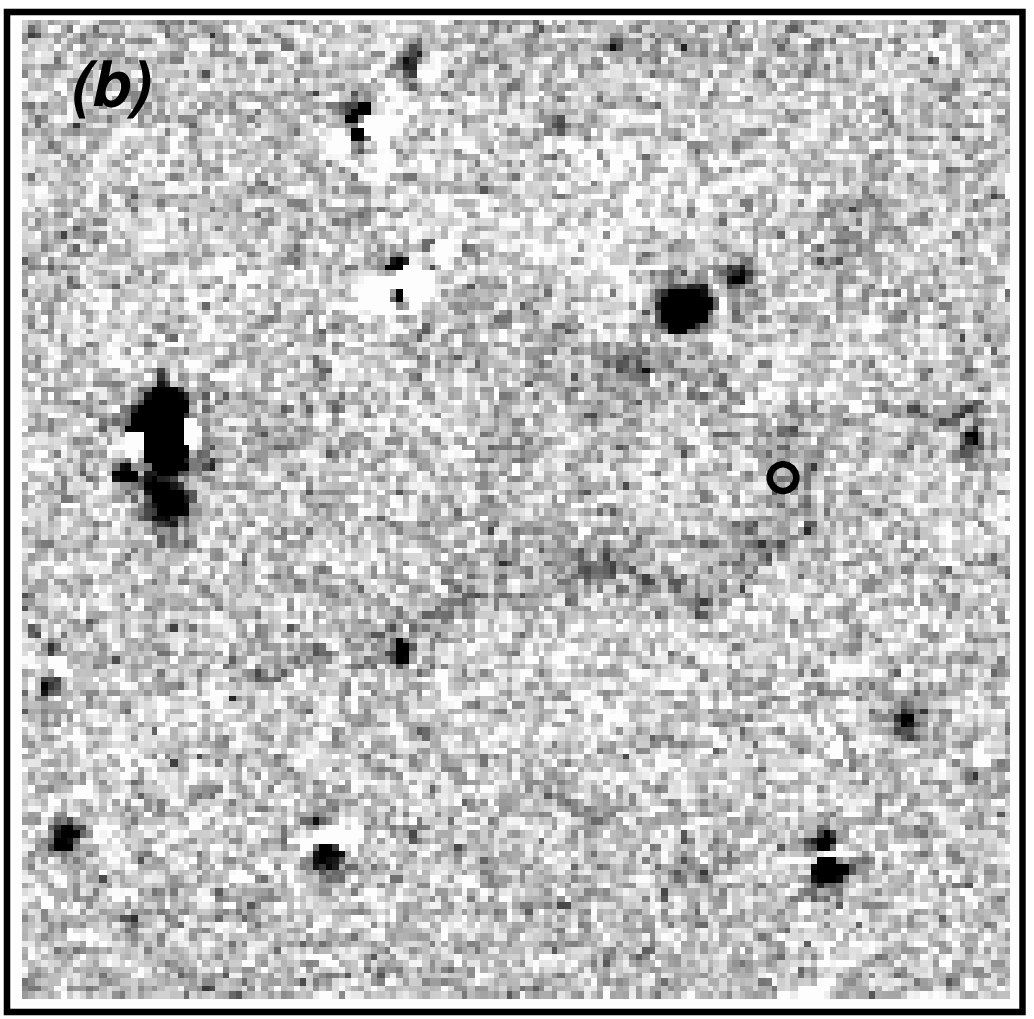}
   \includegraphics[width=6cm]{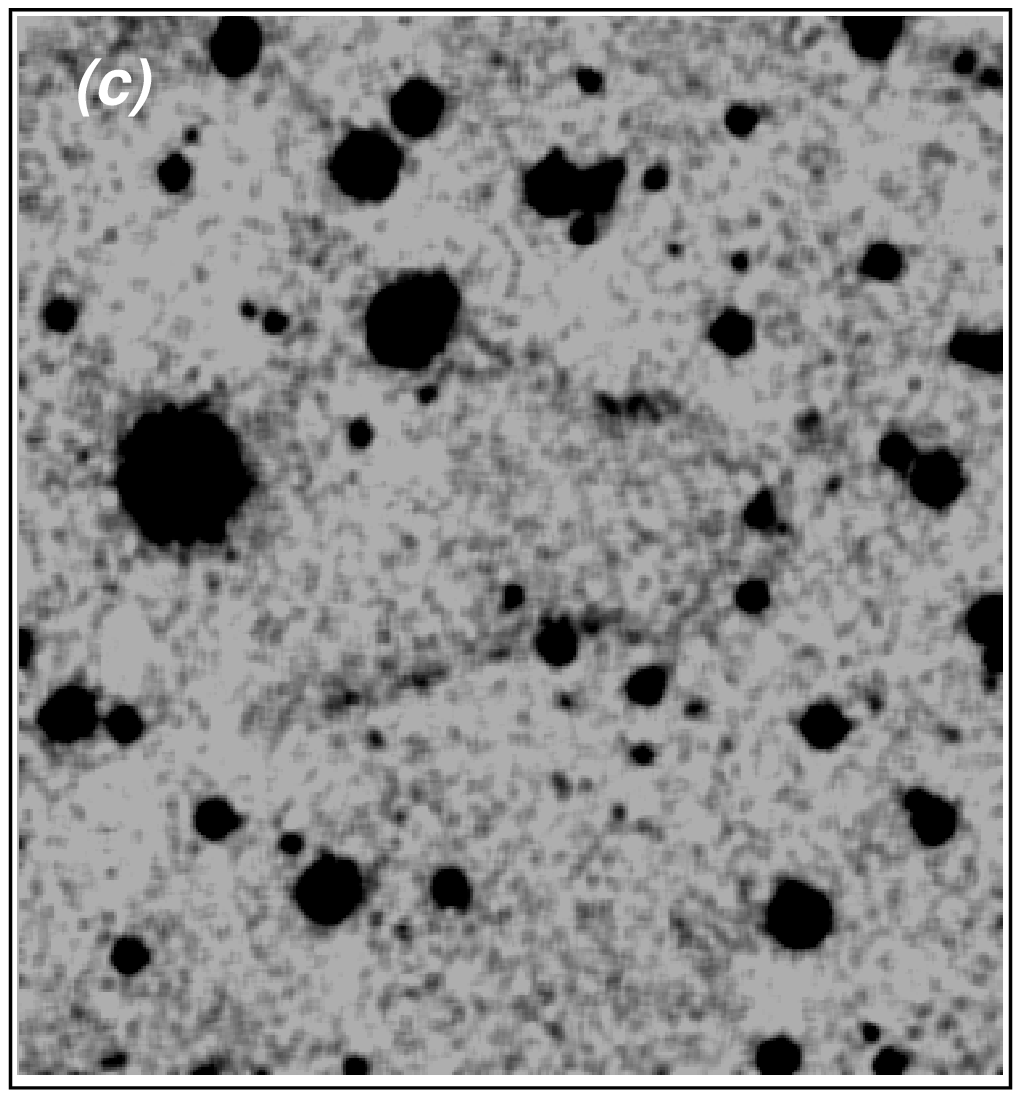}
   \includegraphics[width=6cm]{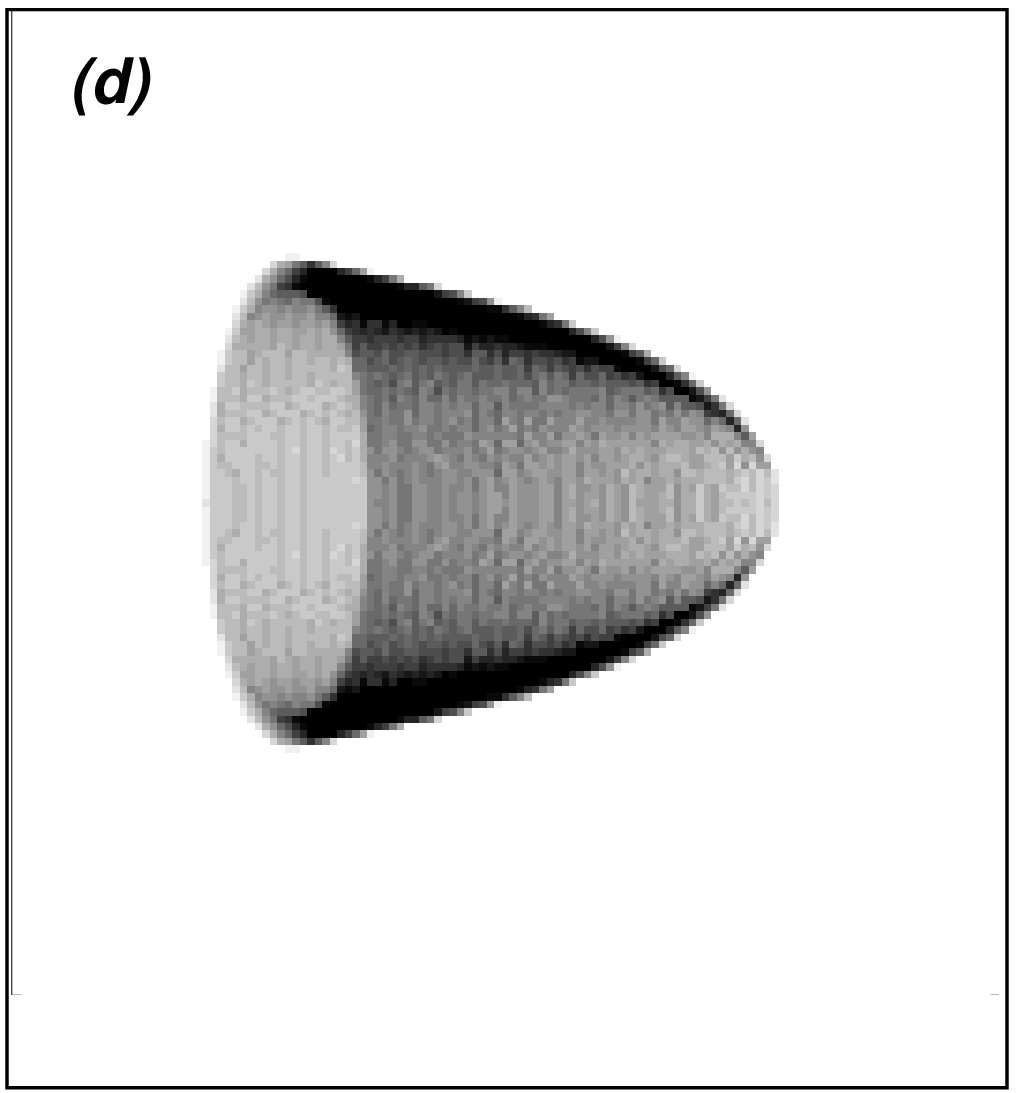}
      \caption{($a$) SUSI-2/NTT discovery image (\Ha\ $+$ continuum) 
of the nebula 
associated with PSR B0740$-$28. The arrow indicates the direction 
and distance travelled by the pulsar over 500 yrs. Coordinate epoch
is J2000.
($b$) Continuum-subtracted \Ha\ image of the
same field taken with TTF on the AAT. Some residuals due to saturated stars
and in-focus ghost images are present. ($c$) Image ($a$) smoothed using a
$5 \times 5 $-pixel boxcar kernel. Panels $a$ --- $c$ are 1 arcmin on a side with north up, 
east left, with the pulsar position indicated by a circle. 
($d$) Illustrative example of a simple surface of rotation 
based on Eqn.~(\ref{chen}), that assumes
the pulsar is moving 20$^{\circ}$ to the plane of the sky.}
   \label{fig1}
   \end{figure*}

The nebula around PSR B0740$-$28 was discovered in narrowband frames
taken through the 656/7~nm filter of SUSI-2 at the 3.5~m New
Technology Telescope (NTT), La Silla (Fig.~\ref{fig1}$a$). A separate
continuum-subtracted narrowband frame (Fig.~\ref{fig1}$b$) 
was derived from subsequent
imaging with the Taurus Tunable Filter (TTF; Bland-Hawthorn \& Jones
1998) at the 3.9~m Anglo-Australian Telescope (AAT). Figure~1($c$) 
shows the NTT image in Fig.~\ref{fig1}($a$) smoothed with a 
$5 \times 5$-pixel boxcar kernel.
The observations are summarised in Table \ref{table2}. Raw frames
were treated using standard techniques for bias and flat-field
correction, image registration and co-addition. CCD defects and cosmic
rays were removed using cross-pixel interpolation on the final frame,
in order to preserve as much of the faint nebula signal as possible.

An astrometric solution was derived for the higher-resolution NTT image
using the {\sc iraf} task {\sl ccmap}. A Legendre polynomial with 
11 coefficients
was found to give the best fit in $x$ and $y$ to 24 USNO Catalog reference 
stars across the field. The fitting option used in {\sc ccmap} 
sees a linear term (with coefficients dependent on translation, scaling, 
rotation and skew), computed separately from a distortion term, which is
a polynomial fit to the residuals of the linear term. 
No distortion residuals were evident in a test transformation of the
USNO star coordinates to those of the image.
The RMS scatter was $(\Delta \alpha, \Delta \delta)$ = 
($\pm 0.028$\as,$\pm 0.038$\as) and uniform across the field.

The location of PSR B0740$-$28 was determined by applying the proper
motion determined by Bailes et al.\ (1990) to the position measured by
Fomalont et al.\ (\cite{fomalont92}). This places the pulsar at
$(\alpha, \delta) = (07$:$42$:$49.041, -28$:$22$:$44.02)$ (J2000) at
the mean time of observation, 1\farcs3 inside the leading edge of the
nebula on its long axis (Fig~\ref{fig1}). It is not coincident with
the prominent star near the head of the nebula.  Bailes et al.\ (1990) 
determine a proper motion position
angle (measured through east) of $\theta = 269\pm1^\circ$, {\em i.e.}
due west.  This is consistent with the direction of the motion implied
by the symmetry axis of the nebula and confirms the association.


\section{Flux and Morphology}

The optical emission from pulsar wind nebulae is purely Balmer.
The \Ha\ flux density in the NTT image was calibrated using
observations of the planetary nebula flux standards 235.3$-$03.9 and
278.6$-$06.7 (Dopita \& Hua \cite{dopita97}).  
Atmospheric extinction was estimated using the CTIO measurements of 
Stone \& Baldwin (\cite{stone83}). For the planetary nebula standards,
the 656/7 filter admits flux from the adjacent \Niib\ and \Niir\ 
lines, in addition to \Ha. The \NII\
contribution for 235.3$-$03.9 (37\%) is more significant than 278.6$-$06.7
(15\%), although the 656/7 fluxes were corrected for it in both cases. 

The \Ha\ flux over the entire PSR B0740$-$28 nebula is $5 \times
10^{-5}$~\photunits, around two orders of magnitude fainter than the
nebulae powered by J0437$-$4715 (Bell et al.\ \cite{bell95}) and
B1957$+$20 (Kulkarni \& Hester \cite{kulkarni88}). 
The \Ha\ images in Fig.~\ref{fig1} show a distinctive key-hole shape
which can be divided into three main regions. The most western
component, or ``head'', has an almost circular shape, with the pulsar
located at the leading edge. Approximately 20\as\ from the front of
the shock, the nebula broadens into a fan-shaped tail and a bridge of
emission connecting the two sides of the nebula can be seen. At the
eastern limit of emission, some 45\as\ from the apex, 
the nebula appears to show a boundary with opposite concavity
to that of the apex, closing the tail.
Figure~1($b$) shows no emission within the interior of the head
although deeper images are needed to confirm this.

Variations in the surface brightness across the extent of the nebula
are seen in Fig.~1$(a)$ and Fig.~1$(b)$; see Table \ref{fluxes} for a
summary. The knot at Region A, located at the juncture of the ``fan''
and ``head'' section, and its southern counterpart both have surface
brightness greater than that of the overall nebula.

\section{Nebular Geometry}

When a star moves at a supersonic velocity with respect to the
interstellar medium (ISM), a bow-shock forms at the interface between
the stellar wind and the ambient medium. For pulsars the wind is
relativistic. Assuming an isotropic wind, the stand-off distance,
$r_{\rm w}$, separating the pulsar from the contact discontinuity of the
shock, can be derived from the balance between ram pressure and the
outflowing wind pressure,
\begin{equation}
r_{\rm w} = \left(\frac{\dot{E}}{4\pi c \rho_0v_{\rm p}^{2}}\right)^{1/2}~{\rm cm}
\label{shock_radius}
\end{equation}
where $\rho_0 = 2.2\times10^{-24}n_{\rm H}$\,g\,cm$^{-3}$ is the ISM
density, $n_{\rm H}$ is the hydrogen number density and $v_{\rm p}$ 
(cm~s$^{-1}$) is the space velocity of the pulsar.

The shock apex was determined by finding the peak emission lying along the
pulsar proper motion direction and is located 1\farcs3 from the
pulsar. However this does not correspond to $r_{\rm w}$, as the Balmer
emission comes from a thin shell upstream from the contact
discontinuity. Aldcroft et al.\ (\cite{aldcroft92}) find for B1957+20
that the measured pulsar-apex separation is 30\% larger than
$r_{\rm w}$. Hence we adopt $r_{\rm w} = $1\farcs0 $= 0.005 d_{\rm kpc}$~pc, 
where
$d_{\rm kpc}$ is the pulsar distance in kpc. The proper motion, when corrected
for galactic rotation, corresponds to a projected space velocity of 
$v_{\rm p} = 1.3\times10^7d_{\rm kpc}/{\rm cos}i$~\cmPerSec, where $i$ is the
angle of this velocity vector to the plane of the sky. Substituting
$v_{\rm p}$ into Eqn.~(\ref{shock_radius}) and rearranging gives:
\begin{equation}
n_{\rm H} = \frac{4{\rm cos}^4i~I_{45}}{d_{\rm kpc}^4}~{\rm cm}^{-3}.
\label{nh}
\end{equation}
Using the upper and lower limits to the distance given in Table
\ref{physParams} and assuming that $i$ is small then $0.002 \simlt
n_{\rm H} \simlt 1$~cm$^{-3}$. If we assume the dispersion measure
distance $d_{\rm kpc} = 2$, (which we argue for in Section~5), then
the density is n$_{\rm H} \sim 0.25$~cm$^{-3}$.

\begin{table} 
\begin{center} 
\caption{Mean Surface Brightness of Different Regions}
\label{fluxes}
\begin{tabular}{cc} 
\hline \hline
Region                                        & Surface Brightness \\
(see Fig.~1$a$)                                 & (erg\,s$^{-1}$\,cm$^{-2}$\,arcsec$^{-2}$) \\
\hline
A   		&  7$\times10^{-19}$ \\
B &  4$\times10^{-19}$  \\
C 		& 5$\times10^{-19}$ \\
entire nebula 					& 3$\times10^{-19}$ \\   
\hline
\end{tabular} 
\end{center} 
\end{table} 

Recent derivations of the geometry of thin momentum-balance bow-shocks
give a generic form for the shock that scales with stand-off distance
(Chen et al.\ \cite{chen96}; Wilkin \cite{wilkin96}).  
We have compared the shape of the PSR B0740$-$28
nebula with the expression of Chen et al.\ (\cite{chen96}),
\begin{equation}
\frac{x}{r_{\rm w}} + 1 = \frac{3}{10}\biggl( \frac{y}{r_{\rm w}} \biggr)^2 + \biggl( \frac{3}{280}\frac{y}{r_{\rm w}} \biggr)^4.
\label{chen}
\end{equation}
Here, our axes lie in the plane of the sky, such that the $y$-axis lies
along the proper motion direction, $x$ is perpendicular to it and the
origin is located at the pulsar.  The best fit to 14
locations around the ``head'' gives a stand-off distance of 
1\farcs2\,$\pm 0.1$ and proper motion position angle of 
$276 \pm 3^{\circ}$ which are consistent
with the measured values. However some deviation from an exact fit
occur on the southern edge of the shock front. Modelling by Wilkin
(\cite{wilkin00}) shows that either wind asymmetry or density
gradients in the ISM are able to reproduce such
variations. Observations of PWNe around other pulsars have
demonstrated marked anisotropies in the wind flow away from the
pulsar, (Hester \cite{hester98}; Gaensler et al.\ \cite{gaensler02}),  
while large variations in the ISM
density are seen in the region of the Guitar nebula (Chatterjee \& Cordes 
\cite{chatterjee02}).


\section{Discussion\label{discussion}}

Cordes (1996) shows that the \Ha\ flux density in
the nebula can be expressed in terms of observables as,

\begin{equation}
F_{\alpha} = \frac{10^{-2.47}Xn_{\rm H}v_7^3\theta_s^2}{{\rm cos}^5i}~{\rm
photons\,cm}^{-2}\,{\rm s}^{-1},
\label{flux}
\end{equation}
where $X$ is the neutral fraction, $v_7 = v_{\rm p}/10^7$ and 
$\theta_{\rm s}$ is
the observed angular distance. The $2$~kpc distance derived from
the dispersion measure (Taylor \& Cordes \cite{taylor93}) is
consistent with the amount of reddening and closed shape of the nebula, in
the following ways.

The observed reddening in the direction of PSR B0740$-$28
is $A_{\rm v} \approx 1.5$ magnitudes at $d \approx 2$~kpc 
and distances beyond,
corresponding to a reduction in \Ha\ flux by a factor of
$\sim 3$ (Neckel \& Klare \cite{neckel80}).
Combining this with a neutral fraction $X = 0.01$, (reasonable
for a warm ionised ISM with the n$_{\rm H} \sim 0.25$~cm$^{-3}$
calculated earlier from Eqn.~\ref{nh} with $d_{\rm kpc} = 2$), 
gives good agreement with the observed flux. 
At larger distances
the density derived from Eqn.~(\ref{nh}) implies the pulsar is in the
hot phase of the ISM where the neutral fraction is very small, thereby
further reducing the predicted flux.

The direction of the nebula with respect to the plane of the sky will also
affect the measured flux. Using the shock morphology formula given in 
Eqn.~(\ref{chen}), we generated a three-dimensional 
model of the shock and then
determined its shape projected on the plane of the sky for various values of
$i$. We find that to best match the observed stand-off distance and opening
angle $i$ must be small ($\simlt 25^{\circ}$) and thus its contribution to
the low flux is small.

The overall morphology of the PSR B0740$-$28 nebula is very similar to the
Guitar nebula where the ``neck'' and ``body'' regions (Cordes et
al. \cite{cordes93}) are analogous to our ``head'' and ``fan'' regions
respectively. However, there are some differences; the head is more circular
than the cylindrical neck of the Guitar nebula and the fan is more
rectilinear than the bulbous body of the Guitar nebula. We note, however,
that both nebulae
show stronger \Ha\ emission in
the transition between the narrow and broad regions.  These could indicate
regions of increased density on sub-parsec scales, 
given that the full extent of the
nebula is $\sim 0.6$~pc if 2~kpc away.  Alternatively, the stronger emission
may indicate the transition between 
different physical regimes of the nebula. 

The Guitar nebula emission shows two distinct bubble regions (Cordes et
al. \cite{cordes93}).  The fainter, inner bubble is analogous to
the ``bridge'' of emission that we see in PSR B0740$-$28. It is possible to
generate these apparently closed structures via projection of a truly open
structure. This can occur if the nebula is slightly inclined, 
as is shown in Fig~1($d$) for a cone assumed to be inclined at 20$^\circ$.

The apparently closed nature of the PSR~B0740--28 nebula suggests that
in the rear region of the nebula, the ram pressure due to the pulsar
motion is no longer important and we could instead be seeing the point
where the pressure in the wind is balanced by the ISM pressure --- it
is possible that the break between the ``head'' and ``tail'' components
of the PWN may correspond to the transition between these two sources
of confinement.  However, this explanation is not consistent with the
entire PWN being embedded in gas of uniform density, since the pulsar's
motion must be supersonic to form a bow shock, but in which case there
can be no pressure balance at the rear of the PWN.  
In order to explain the observed morphology, we propose that the pulsar
has recently passed through a strong density gradient in the ISM, such
that the pulsar's motion is supersonic in the comparatively dense
medium ($n_{\rm H} \sim 0.25$~cm$^{-3}$) through which it is currently
propagating, but was subsonic in low density material ($n_{\rm H} \la
0.003$~cm$^{-3}$) surrounding the closed end of the PWN.
Similar arguments can be invoked to explain the morphology
of the Guitar PWN powered by PSR~B2224+65.

Considering the latter region, the requirement that the pulsar be
moving slower that the local sound speed implies an upper limit to the
pulsar's distance of 3--4~kpc, as no ISM component has a sound speed
faster than $\sim500$~km~s$^{-1}$. If PSR~B0740--28 is at 2~kpc, then
the pulsar wind pressure, $P_{\rm w} = \dot{E}/(4\pi r_b^2c)$ approximately
balances the ISM pressure, $P_{ISM} \sim 5\times10^{-14}$~Pa, at the
the location of the closed region (where $r_{\rm b}$ is the distance to the
back of the nebula), providing further support for the lower distance
argued for above.  In the interpretation we propose, the closed region
of the PWN corresponds to the static or ``ghost'' nebulae proposed  by
Blandford et al.\ (\cite{blandford73}). The faintness of this
region is consistent with the expectation that such sources should
generally never be observable (Gaensler et al.\ \cite{gaensler00}); 
in the case of PSR~B0740--28, it was only through the presence of
brighter emission near the pulsar that this closed region was
identified.

Bucciantini \& Bandiera (\cite{bucciantini01}) have recently shown
that classical thin-shock models are not appropriate for pulsar
bow-shock nebulae. However, here we find reasonable agreement between
the classical models and the shape of the head of the
PSR B0740$-$28 nebula.
It is likely though that the microphysical processes they
discuss are responsible for at least some of the morphological
characteristics of these PWNe. Indeed Bucciantini \& Bandiera
(\cite{bucciantini01}) point out that perhaps PSR B0740$-$28 is an
intermediate case between the nebulae associated with the MSPs and the
Guitar nebula. Interestingly, they also predict that, based on the ISM
conditions, this nebula should be readily detectable.
However, we find that it is considerably less-so than the others.
Clearly, increasing the presently small sample of optical bow-shock nebulae
remains a key step towards being able to generalise about the ISM conditions
within the vicinity of pulsars.

\begin{acknowledgements}
We are grateful to the NTT Team and J.~Bland-Hawthorn for the quality
of the data obtained at the NTT and AAT respectively. We are also grateful
to the referee, R.~Bandiera, whose comments helped to improve the final 
draft. BWS is supported by NWO Spinoza
grant 08-0 to E.~P.~J.~van den Heuvel. BMG acknowledges the support of a CfA
Clay Fellowship.  DHJ would like to acknowledge R.~A.~Faulkner (1945--2001)
for encouraging an early interest in astronomy.
\end{acknowledgements}


\end{document}